\pdfoutput=1
\documentclass[aps,prb,floatfix,twocolumn,footinbib]{revtex4-1} 
\usepackage{epsfig}
\usepackage{hyperref}
\usepackage{cancel} 
\usepackage{amsmath, amssymb} 
\usepackage{graphicx}
\usepackage{epsf}
\usepackage{braket}
\usepackage{appendix}
\usepackage{color}

\graphicspath{{Figures/}}

\usepackage{epsfig}
\usepackage{hyperref}
\usepackage{cancel} 
\usepackage{amsmath, amssymb} 
\usepackage{graphicx}
\usepackage{epsf}
\usepackage{braket}
\usepackage{appendix}
\usepackage{color}
\usepackage{todonotes}
\usepackage{amssymb}
\usepackage{gensymb }
\usepackage[pagewise]{lineno}
\usepackage[normalem]{ulem}

\begin{document}
\title{Dark-field imaging as a non-invasive method for characterization of whispering gallery modes in microdisk cavities}

\author{D.\,A.~Baranov$^1$}
\author{K.\,B.~Samusev$^{1,2}$}
\author{I.\,I.~Shishkin$^{1,2}$}
\author{A.\,K.~Samusev$^{1,2}$}
\author{P.\,A.~Belov$^{1}$}
\author{A.\,A.~Bogdanov$^{1,2}$}
\email{bogdan@metalab.ifmo.ru}

\affiliation{
$^{1}$ ITMO University, 197101 St.~Petersburg, Russian Federation \\
$^{2}$Ioffe Institute, 194021 St.~Petersburg, Russian Federation}

\date{\today}



\begin{abstract}
{Whispering gallery mode microdisk cavities fabricated by direct laser writing are studied using dark-field imaging and spectroscopy in the visible spectral range.  {Dark-field imaging allows us to directly visualize the spatial intensity distribution of whispering gallery modes. We extract their azimuthal and radial mode indices from dark-field images, and find the axial mode number from the dispersion relation.}
The scattering spectrum obtained in the confocal arrangement provides information on the density of optical states in the resonator. The proposed technique is a simple non-invasive way to characterize the optical properties of microdisk cavities.}
\end{abstract}



\maketitle

	

Microcavities with circular symmetry can support so-called whispering gallery modes (WGMs) formed by the total reflection of the electromagnetic wave from the side wall of the cavity. WGM microcavities can have huge $Q$-factors, reaching a magnitude of up to 10$^{10}$ [\onlinecite{righini2011whispering}].  {High $Q$-factors allow to accumulate the electromagnetic energy density sufficient for a considerable manifestation of nonlinear effects~[\onlinecite{righini2011whispering}].} Miniaturization of the cavity sizes results in the exponential increase of the radiation losses which can be suppressed by means of auxiliary plasmonic structures~[\onlinecite {righini2011whispering}].
WGM resonators are widely used in sensors and detectors because of their high $Q$-factor and a narrow resonance line width providing high spectral sensitivity to the environment (see, e.g., Ref.~[\onlinecite {he2012whispering,lahoz2012whispering,stock2013chip, latawiec2015chip}]). Even traces of single molecules or nanoparticles can be efficiently detected by means of the spectral shift of WGM resonances~[\onlinecite {he2011detecting,chen2012optomechanical,kim2014surface}]. WGM resonators are also applicable for lasing, filtering, switching, and modulation of optical signals  [\onlinecite{Ilchenko2006,maleki2015tunable, matsko2012multi,Bogdanov2015,Kryzhanovskaya2014,Maximov2014}].

The full electromagnetic description of microcavities contains the information about the spectrum (eigenvalues) and field distribution of the modes (eigenfunctions), which is enough to build Green function of microcavity~[\onlinecite {novotny2012principles}]. Eigenfrequencies and $Q$-factors can be estimated experimentally, for example, from the transmission spectrum of a fiber~[\onlinecite {Lin2010}] or a prism~[\onlinecite {Schunk2014}] coupled to the resonator or from the spectrum of a photoluminescent  material placed inside the resonator~[\onlinecite {righini2011whispering, PhysRevA.76.023816}].  { The latter method demands special procedures during the fabrication process. Spectrum measurements with a fiber or a prism coupled to the waveguide is quite demanding because of the exponential dependance of the coupling coefficient on the distance between the cavity and the fiber. In the case of under-coupling regime, the resonance in transmission  spectrum can be very weak and not detectable. In the over-coupling regime, the fiber distorts the spectrum of the cavity. Moreover, the coupling coefficient depends on the mode indices of WGM. }

The field distribution of the eigenmodes can be measured using near-field scanning optical microscopy (NSOM)~[\onlinecite {Balistreri1999,Klunder2000,Gotzinger2001,Schmidt2012}].  {This method as well as a fiber coupled to a cavity is also invasive}.  A probe tip of a NSOM distorts spectrum, $Q$-factor, and field distribution of modes, especially in high-Q cavities. Therefore, an additional treatment of the measured data is needed.  Sometimes, it is a quite tricky and nontrivial problem~[\onlinecite{le2014simultaneous, permyakov2015probing}]. In this sense, non-invasive methods, i.e. those which do not distort the properties of the resonator, become more relevant and topical.    

\begin{figure}\centering
	\includegraphics[width=1\linewidth]{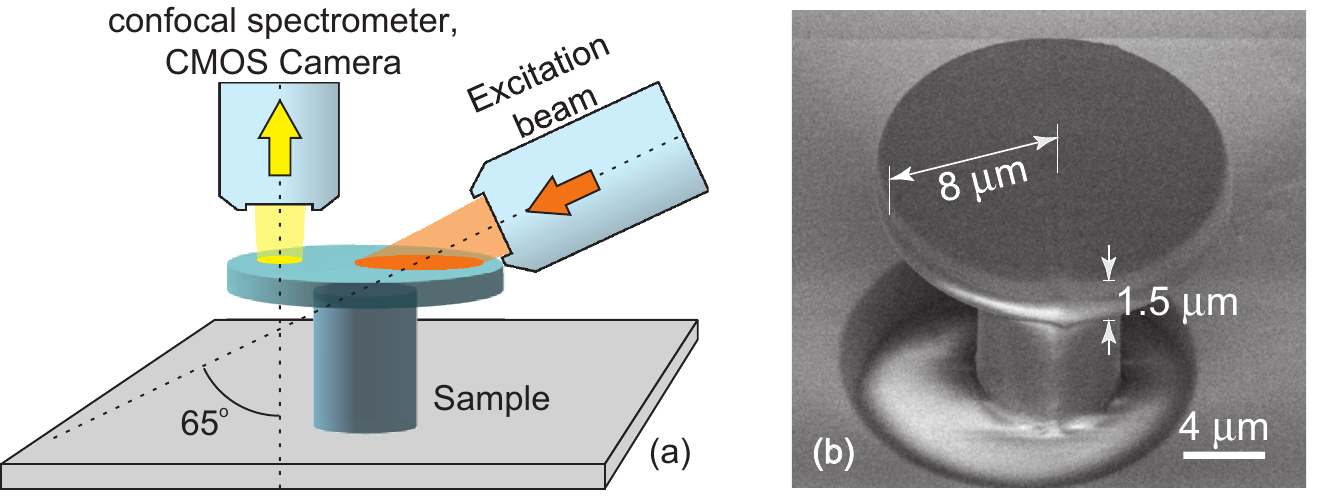}		
	\caption{\label{fig:sem} (a)~Experimental setup for dark-field imaging and spectroscopy of WGM microcavities. The sample is excited with white light from the side at oblique incidence. The scattered light is collected from the top by an objective with NA~=~0.55. (b)~SEM image of the fabricated microdisk. The radius of the disk $R=8~\mu$m, disk thickness $H = 1.5~\mu$m. }
\end{figure}

In this work, we demonstrate a novel non-invasive technique of microcavity  {characterization based} on dark-field imaging and spectroscopy.  {This method uses} different optical channels for excitation beam and collection of the radiation scattered by the microcavity~[see Fig.~\ref{fig:sem}(a)].  {The coupling of the obliquely incident Gauss beam with WGMs occurs due to the radiation losses mainly coming from the sidewall roughness. The proposed technique allows to directly observe the intensity profile of WGMs without the need to couple WGM resonators to a fiber or a prism or to embed dye molecules inside the cavity.}  {Note that in Ref.~[\onlinecite{Dong2008}] a similar imaging technique of WGMs was developed. However, in contrast to the method proposed here, it applied only to deformed cavities and does not allow to extract azimuthal mode index.}

The fabrication of WGM microresonators was performed with negative-tone photoresist based on hybrid organic-inorganic material based on zirconium propoxide with Irgacure 369 as photoinitiator [\onlinecite{Ovsianikov2008}]. The material exhibits low shrinkage upon polymerisation, which guarantees good correspondence of the fabricated structure to suggested model. A commercial direct laser writing (DLW) system (Lazer Zentrum Hannover, Germany) equipped with Ti:Sa oscillator (TiF-100, Avesta Project, Russia) seeding 100-fs pulses centered around 790 nm wavelength with the repetition frequency of 80 MHz was used for the exposure of photoresist. The laser beam was focused in a volume of the photoresist through a glass coverslip by 100$\times$ immersion objective with numerical aperture NA = $1.4$ (Carl Zeiss, Germany). The positioning of the sample was performed by motorized linear air-bearing translators (Aerotech inc., USA). The samples were prepared by drop-casting the photoresist on 170~$\mu$m thick glass coverslips used as substrates and pre-baked at 70$^{\circ}$ for 1 hour. After the exposure the samples were developed in isopropanol for 30 minutes.


The  mode intensity distribution and scattering spectrum of the microdisk were analyzed using a home-built dark-field microscope with independent excitation (side) and collection (upper) optical channels [see Fig.~\ref{fig:sem}(a)].  {The fabricated} microresonators were illuminated by a supercontinuum source (Fianium Superchrome) with a tunable band-pass filter with minimal band of $5$~nm. TM-polarized light was focused with an objective (Mitutoyo MPlanApo 10$\times$, NA~=~0.28) on the sample surface at oblique incidence of $65^\circ$. The light scattered by a single microdisk was collected with the second objective (Mitutoyo MPlanApo 50$\times$, NA~=~0.55) to observe the spatial distribution of scattered light. Experimental spectra were acquired using Horiba LabRAM HR confocal spectrometer from the spot located at the edge of the disc as shown in Fig.~\ref{fig:sem}(a).

\begin{figure}[h]
	\centering
	\includegraphics[width=0.9\linewidth]{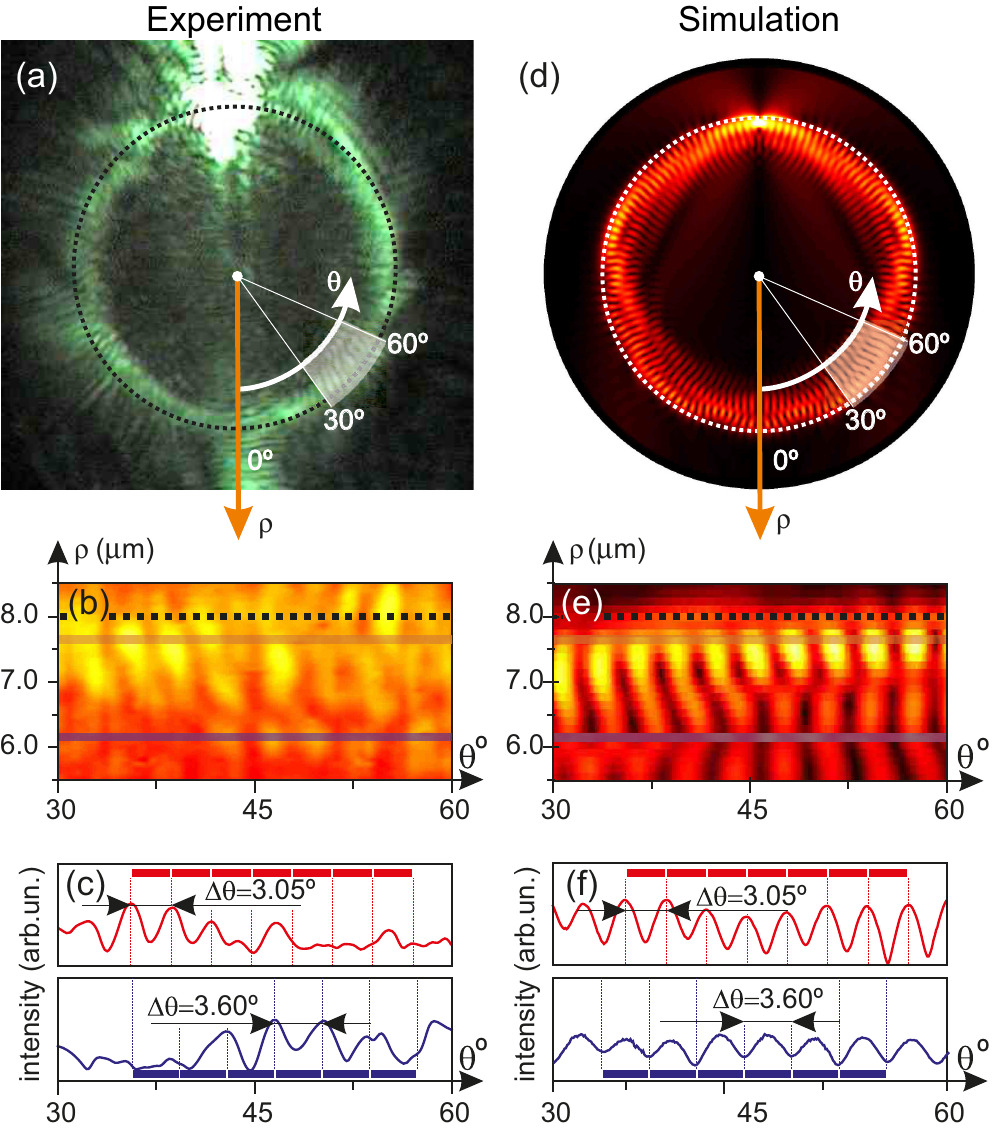}
	\caption{\label{fig:peakmap} (a) Experimental dark-field image of a microdisk resonator obtained at  $\lambda~=~525$~nm excitation.  {The bright region on the top of the image is the excitation spot.} (b) A developed sector $30^\circ-60^\circ$ of an experimental image represented in angle-radius coordinates. (c) Intensity cross-sections of the development (b) obtained at $r=7.7~\mu$m (red line) and  $r=6.2~\mu$m (blue line). (d-f) Numerically simulated electric field intensity distribution (d), its respective angle-radius development (e) and cross-sections (f). Dotted lines in (a, b) and (d, e) indicate microdisk radius $R=~8~\mu$m. } 
\end{figure}

Dark-field image of the microcavity obtained under oblique  {excitation} with $\lambda=525$~nm is shown in Fig~\ref{fig:peakmap}~(a).  {One can clearly see a periodic distribution of the intensity along the microdisk's edge which is the manifestation of the excited WGMs.} 
 {The incident beam lies in the symmetry plane of the microcavity. It results in equiprobable excitation of the clock- and anti-clockwise WGMs which form the observed standing wave.} Because of the relatively large size of the cavity its spectrum is  {quite dense and several modes are excited simultaneously.} Each of them can be described by azimuthal ($m$), radial ($r$), and axial ($z$) mode numbers which have clear physical meaning -- numbers of  {mode field maxima} in the corresponding directions.

Detailed view of the intensity pattern is shown in Fig.~\ref{fig:peakmap}~(b). It is obtained with conformal mapping of the hatched area from Fig.~\ref{fig:peakmap}~(a). Two series of the intensity maxima at different distances from the center of the cavity correspond to the fundamental ($r=1$) and higher-order radial mode  {($r>1$)}. The angular dependence of WGM field is described through a factor $\exp(\pm i m \theta)$, where $\theta$ is the azimuthal angle. Using fast Fourier transform we obtain angular distances $\Delta \theta$ between  {the neighbouring} maxima of intensity for each mode [see Fig.~\ref{fig:peakmap}~(c)] and find azimuthal mode number as $m = 180^\circ / \Delta \theta$. For the modes under consideration we obtain $m=59$ and $m=50$, respectively. From the condition of the spectral proximity of these modes we  find that for $m=50$ the radial index is $r=3$. Thus, we  {directly} identified azimuthal and radial mode numbers as $m=59, r=1$ and $m=50, r=3$.

Axial mode number $z$ can be obtained from approximate relation between the resonance wavelength $\lambda$ and mode indices~$m, r, z$~[\onlinecite {gorodetskii2007eigenfrequencies}]:

\begin{equation} \label{TM}
\lambda=\frac{2\pi R n}{\gamma_{m,r}\sqrt{1+\left(\frac{z\pi R}{h \gamma_{m,r}}\right)^2}}=\frac{2\pi R n_\text{e}}{\gamma_{m,r}}.
\end{equation}
Here, $R$ and $h$ are the radius and the height of the microdisk, $\gamma_{m,r}$ is the $r^\text{th}$ root of Bessel function of  $m^\text{th}$ order, $n$ and $n_\text e$ are refractive index of the microdisk material and effective refractive index of the mode, respectively. The extracted mode indices and $n_\text e$ for different excitation wavelengths are shown in Table~\ref{modeTable}.

\begin{figure}\centering
		\includegraphics[width=0.8\linewidth]{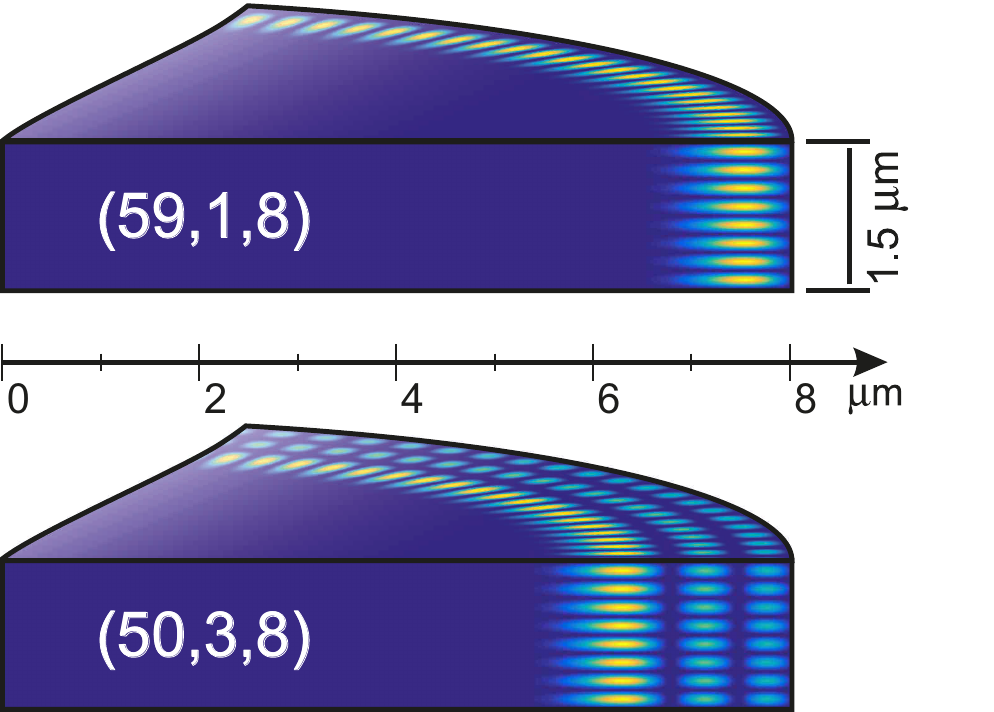}		
		\caption{\label{fig:pizza}  Intensity distribution of WGMs with $m=59$, $r=1$, $z=8$ and $m=50$, $r=3$, $z=8$ calculated using the approach from Ref.~[\onlinecite{Frateschi1996}].}
	\end{figure}

In order to verify the equity of the proposed procedure we carried out 2D numerical simulation of the dark-field imaging experiment using Comsol Multiphysics. The excitation beam is substituted with an electric dipole oriented  {normally to the face of the disk}. The thickness of the microdisk is taken into account through the effective  {refractive index of the mode} $n_\text e$. The simulated intensity distribution [Fig.~\ref{fig:peakmap}~(d-f)] demonstrates good qualitative agreement with the experimental image [Fig.~\ref{fig:peakmap}~(a-c)]. The verification of the extracted radial mode indices is provided through the comparison of the experimental mode maximum position [Fig.~\ref{fig:peakmap}~(b)] and the result of 3D simulation (Fig.~\ref{fig:pizza}) obtained using the approach from Ref.~[\onlinecite {Frateschi1996}]. 



\begin{table}[htbp]
 	\centering
 	\caption{\bf Whispering gallery modes observed experimentally at different excitation wavelengths.}
 	\label{modeTable}
 	\begin{tabular}{ccc}
 		\hline
 		$\lambda$&  Azimuthal, radial \& axial &   $n_{\text e}$\\
 		(nm) &   indices ($m$, $r$, $z$) &  \\
 		\hline
 		
 		525 & (59, 1, 8)   & $0.694$\\
 		525 & (50, 3, 8)   & $0.707$\\
 		550 & (61, 1, 7)   & $0.750$\\
 		550 & (53, 3, 7)   & $0.777$\\
 		575 & (64, 1, 7)   & $0.820$\\
 		575 & (58, 3, 7)   & $0.874$\\
 		
 	\end{tabular}
 	
 \end{table}




 {The physical meaning of $n_\text e$ is the ratio of the wavelength in a plane wave (in vacuum) and in a WGM at the same frequency. Value of $n_\text e$ becomes less than unity only for high-order axial modes. This fact is exploited for the simulation of $\varepsilon$-near zero metamaterials using hollow metal waveguides [\onlinecite {pacheco2014e}].

According to the Abbe limit [\onlinecite {novotny2012principles}], we should be able to resolve only the neighbouring intensity maxima placed at a distance bigger than $1.22\lambda/(2\text{NA})$. 
For WGMs it is equivalent to the relation
$n_\text e \leq {\text{NA}}/{1.22}=0.45.$
However, in the experiment, we have the resolution overcoming the Abbe limit (see Table~\ref{modeTable}). This contradiction is explained by the fact that neighbouring intensity maxima of WGMs represent anti-phase coherent sources whose image always consists of distinct maxima regardless of the distance between the sources~[\onlinecite {Nayyar1975}]. Therefore, fundamentally, two anti-phase coherent sources are always resolvable in a noise-free environment~[\onlinecite{Cesini1979}]. In practice, the resolution is limited by sensitivity of the detector and background noise level. In our case, the main source of background intensity is diffuse scattering from the top face of the cavity.

}

\begin{figure}\centering
	\includegraphics[width=1\linewidth]{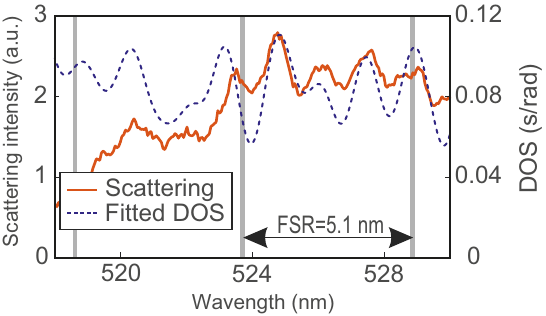}		
	\caption{\label{fig:DOSandScatter} Experimental dark-field spectrum (red line) and density of optical states (blue dashed line). Vertical grey lines indicate fundamental modes, i.e. the modes with first radial and axial indices.}
\end{figure}

 {To gain further insight into the optical properties of the cavity, we carried out of the spectral measurement of the cavity. In order to minimize the diffuse background  we  measured the microcavity spectrum from the spot with a diameter $\approx$1.5~$\mu$m located at the edge of the cavity [$\theta=90\degree$, see Fig.~\ref{fig:sem}(a)].} Experimental scattering spectrum of microdisk is shown in Fig.~\ref{fig:DOSandScatter}.  {One can see that it has an oscillation behaviour. The oscillations can be explained by low finesse of the cavity spectrum consisting of many overlapped resonances. The oscillation period of the scattering spectrum is about 2~nm. It is less than free spectral range (FSR) $\Delta\lambda=5.1$~nm, i.e. mode spacing between the two fundamental WGMs with neighbouring azimuthal numbers. It is a manifestation of higher-order radial mode.} 

 {With the assumption that excitation probability of WGMs has weak dependance on the mode number,  the scattering intensity spectrum is proportional to the density of optical states (DOS) $\rho(\omega)$, which is one of the  most fundamental quantity in nano-optics~[\onlinecite {novotny2012principles}].} In particular, in the case of infinite $Q$-factors, DOS is given by $\rho(\omega)=\sum_{s}\delta(\omega~-~\omega_{s})$. If the $Q$-factor is finite, the Dirac delta function  {$\delta(\omega-\omega_s)$} should be replaced by the Gaussian distribution:

	
\begin{equation}
	\rho(\omega)=\frac{1}{\sqrt{2\pi}}\sum_{s}\frac{Q_s}{\omega_s}\exp\left(-Q_{s}^2\frac{(\omega-\omega_s)^2}{2\omega_s^2}\right).
	\label{eq:DOS}
\end{equation}
Here, $\omega_s$ and $Q_s$ are the eigenfrequency and $Q$-factor of the mode with index $s$. The eigenfrequencies $\omega_s$ were found from full-wave numerical simulation using Comsol Multiphysics software and approach developed in Ref.~[\onlinecite {oxborrow2007traceable}]. This approach was approved in the previous studies (see Ref.~[\onlinecite{Bogdanov2015}]).   For large cavities, the $Q$-factor is mainly  {limited by the scattering from surface roughness}, which is defined by roughness root mean square $\sigma$ and correlation length of surface roughness profile $L$~[\onlinecite {gorodetskii2007eigenfrequencies, borselli2005beyond}]. For the sample under consideration, atomic force microscopy (AFM) measurements and subsequent treatment yields  $\sigma=12$~nm, $L_\text{c}=80$~nm, and $Q\approx2\times10^3$.  It corresponds to a spectral broadening of $0.34$~nm for $\lambda=525$~nm. 
Roughness parameters are governed by both technical details of DLW setup (output laser power stability, voxel size etc.) and by the fabrication procedure~[\onlinecite{DLW_fabrication}].


The calculated DOS [see Eq.~(\ref{eq:DOS})] is shown in Fig.~\ref{fig:DOSandScatter}. One can see that DOS and spectrum of the scattering intensity  have similar oscillation  {period. It confirms the correctness of Q-factor value estimated from the AFM measurements.}  {Breaking of the proportionality between frequency dependence of scattering intensity and density of optical states can be induced by diffuse scattering, finite size of the collection spot, finite aperture of objective in the collection channel, and by a dependence of WGM excitation probability on the mode number.}

In conclusion, by the example of whispering gallery mode microdisk cavities fabricated by direct laser writing, we have demonstrated that the combination of dark-field imaging and spectroscopy provides non-invasive measurement of mode spatial distribution, spectrum of eigenmodes, mode indices, and density of optical states. Such technique can prove extremely useful in the case high quality factor cavities which are very sensitive to environment and, therefore, demand non-invasive measurements.

The authors acknowledge Ivan Sinev for fruitful discussion and proofreading.

\acknowledgements

The technological part of this work was supported by the Russian Foundation
for Basic Research (grant \#14-29-1017, grant \#16-32-00566, grant \#15-57-45141) and the experimental part was supported by the
Russian Science Foundation (grant \#15-12-00040).

\bibliography{DLW-1}

\end{document}